\documentclass[12pt,a4paper]{article}

\usepackage[margin=1in]{geometry}
\usepackage[doublespacing]{setspace}

\usepackage{amsmath}
\usepackage{amssymb}
\usepackage{amsthm}
\usepackage{mathtools}
\usepackage{mathrsfs}
\usepackage{bm}

\usepackage{graphicx}
\usepackage{booktabs}
\usepackage{multirow}
\usepackage{subcaption}
\usepackage{rotating}

\usepackage{enumitem}
\usepackage{algorithm}
\usepackage{algpseudocode}

\usepackage{xcolor}
\usepackage{url}
\usepackage[hidelinks]{hyperref}

\usepackage[authoryear,round]{natbib}

\usepackage{appendix}

\graphicspath{{Fig/}}

\theoremstyle{plain}
\newtheorem{theorem}{Theorem}
\newtheorem{proposition}[theorem]{Proposition}
\newtheorem{lemma}{Lemma}

\theoremstyle{definition}

\newtheorem{assumption}{Assumption}
\newtheorem{condition}{Condition}

\theoremstyle{remark}
\newtheorem{remark}{Remark}

\newcommand{\mcal}{\mathcal}
\newcommand{\mbb}{\mathbb}
\newcommand{\mbf}{\mathbf}
\newcommand{\T}{\top}

\DeclareMathOperator{\rank}{rank}
\newcommand{\mscr}{\mathscr}
\DeclareMathOperator{\KL}{KL}
\DeclareMathOperator*{\argmin}{arg\,min}
\DeclareMathOperator{\PL}{PL}
\DeclareMathOperator{\VR}{VR}

\newcommand{\R}{\mathcal{R}}
\newcommand{\PP}{\mathbb{P}}

\title{\textbf{Population-Level Generative Modeling for Ranking Data}}

\author{
Zhaoyang Shi\thanks{Corresponding author: 
\href{mailto:zyshi10m@outlook.com}{zyshi10m@outlook.com}}\\[0.5em]
Center for Applied Mathematics, Fudan University\\
Shanghai, China
}

\date{}

\begin{document}

\maketitle



\begin{abstract}
Ranking data arise in scientific and machine learning applications, including recommendation systems, information retrieval, voting, marketing, and AI preference ranking from human feedback. Existing statistical work has primarily focused on inference tasks such as preference estimation, rank aggregation, and ranking prediction. However, generating realistic synthetic rankings from an observed population is important for privacy-preserving data sharing, benchmark construction, simulation, and uncertainty quantification. This task is challenging because rankings are high-dimensional combinatorial objects with non-Euclidean dependence structures, while ranking populations often exhibit substantial preference heterogeneity. We propose a framework for population-level generative modeling through a latent preference simplex embedding. It estimates a low-dimensional latent preference simplex through a likelihood-based ranking model, leverages flow matching to learn the population distribution of latent preferences, and generates new rankings through the fitted probabilistic ranking model. We show that ranking generation admits an oracle reduction to latent distribution learning and derive finite-sample generative guarantees that clarify how the number of items, ranking length, and latent dimension affect accuracy. Experiments on synthetic and real datasets demonstrate improved population-level fidelity and provide a statistically interpretable representation of preference heterogeneity.
\end{abstract}


\section{Introduction}
Ranking data are ubiquitous in modern statistical learning and data science. They arise naturally in recommendation systems \citep{liu2025preference,chen2024softmax}, information retrieval \citep{fang2024scaling,chowdhury2025rankshap}, online advertising \citep{qiu2025unirom,yang2024parallel}, preference elicitation \citep{mukherjee2024optimal,li2025eliciting}, voting \citep{hosseini2024surprising}, marketing \citep{yinexpress}, and, more recently, AI preference ranking from human feedback \citep{xiao2025algorithmic,shi2025fundamental}, an increasingly important component of large language model alignment in which human evaluators rank alternative model outputs according to preference. Unlike scalar labels or isolated pairwise comparisons, ranking data encode relative preferences among multiple competing items within a single observation and therefore provide a richer description of comparative judgment. This structured and highly informative nature makes ranking data central to a broad range of statistical tasks, including preference estimation, consensus analysis, heterogeneity modeling, and prediction. Consequently, the statistical modeling and analysis of ranking data have attracted sustained interest over the past several decades.

Among the broad range of applications and sustained methodological development, the statistical literature on ranking data has primarily focused on inference problems. Existing work has largely been devoted to estimating latent preference scores \citep{fan2025ranking,dong2026statistical}, recovering consensus or population rankings \citep{zhu2023partition,chen2022optimal}, modeling heterogeneous preference structures \citep{mao2022learning,pearce2025modeling,caron2014bayesian}, learning ranking models from pairwise or partial rankings \citep{chen2019spectral,han2025unified}, predicting unobserved or future rankings and quantifying the associated uncertainty \citep{dai2021scalable,duchi2013asymptotics}. Classical probabilistic models, including the Plackett–Luce, Bradley–Terry, and Mallows models together with their numerous extensions, provide the foundation for these inference tasks.

Despite these developments, a fundamentally different statistical problem has received comparatively little attention: \emph{population-level generative modeling for ranking data}. Rather than performing an inferential task such as estimating latent preferences or predicting rankings, the objective is to learn the population-level mechanism underlying ranking behavior and subsequently generate new synthetic rankings from the learned distribution. Such a capability is increasingly important for privacy-preserving ranking data synthesis, benchmark construction for evaluating ranking algorithms, uncertainty quantification through synthetic replications, simulation of complex decision systems, and scalable preference generation for downstream training, evaluating, and aligning AI systems. 

Formally, consider a collection of \(n\) items and a population of $m$ rankers. For each ranker
\(j=1,\ldots,m\), we observe a complete or partial ranking of their preferences over $n$ items:
\(
R_j=(i_{j1}\succ i_{j2}\succ\cdots\succ i_{jL_j}),
\)
where $1\le L_j\le n$,
\(
i_{j\ell}\in\{1,\ldots,n\},
\)
and the ranked items are distinct. The case \(L_j=n\) corresponds to a complete ranking, while \(L_j<n\) corresponds to a top-\(L_j\) partial ranking. The observed ranking data are
\(
\mathcal R=\{R_1,\ldots,R_m\}.
\)
Throughout, the ranking lengths \(\{L_j\}_{j=1}^m\) are treated as design quantities. The rankings are viewed as independent realizations from an
unknown population.
Given only
\(
\{R_1,\ldots,R_m\},
\)
our objective is to learn this population and construct a generative model that,
for a prescribed ranking length \(L_{\rm gen}\), generates new synthetic rankings
\(
\widetilde R_1,\widetilde R_2,\ldots
\)
from the learned distribution. This problem is more challenging than generative modeling for ordinary Euclidean data. First, rankings are high-dimensional combinatorial objects with strong structural dependence. A complete ranking of \(n\) items lies in a permutation space of cardinality \(n!\), while a top-\(L\) ranking belongs to a space of size \(n!/(n-L)!\), making direct distributional learning statistically and computationally difficult. Second, changing the position of one item necessarily alters its ordering relative to multiple others, meaning ranking coordinates have complex dependence and cannot be modeled independently. Third, real ranking populations are also highly heterogeneous, with individuals exhibiting distinct preference patterns. These challenges therefore leave a central statistical question unresolved:
\begin{center}
\emph{How can we develop accurate generative models for ranking data that preserve the underlying ranking structure while remaining statistically interpretable?}
\end{center}

\subsection{Latent preference simplex embedding with flow matching}
Our central idea (see details in Section \ref{sec:ourmethod}) is to represent high-dimensional and heterogeneous ranking behavior through a low-dimensional and statistically interpretable preference simplex. Under the Plackett-Luce model, each ranker \(j\) is associated with a utility vector \(q_j\) that determines the distribution of the observed ranking. In many real-world ranking settings, rankers exhibit a collection of recurring preference types but weigh them differently when ranking the items. This observation naturally motivates a simplex representation of the utility $q_j$ in which the vertices stand for latent utilities corresponding to those preference types and a ranker combines these latent utilities with different preference weights to form their utility $q_j$. We embed the observed rankings into this low-dimensional latent preference simplex by estimating latent utilities and preference weights, use flow matching to learn the population distribution of these weights to generate new preference weights, and decode them into rankings through the fitted Plackett-Luce model.

The proposed framework possesses several attractive properties. First, it reduces generation from a high-dimensional ranking space to a low-dimensional preference simplex. Second, the latent preference simplex provides a direct statistical interpretation in terms of ranker preferences. Third, the framework separates statistical estimation from generative modeling, allowing different modern generative models to be incorporated and thereby providing a flexible and scalable approach. 

\subsection{Related work}
A large statistical literature on ranking data focuses on estimating preference parameters, recovering latent rankings, aggregating
rankings, predicting unobserved comparisons, and quantifying uncertainty. Representative
examples are discussed in the Introduction. However, their inferential target is different and they cannot address the joint problem of learning the population distribution underlying observed rankings and generating new rankings from the learned distribution.

Mixture extensions of ranking models accommodate population heterogeneity
through latent components, with each ranker generated from one of several
component-specific ranking models
\citep{murphy2003mixtures,gormley2008mixture,mollica2017bayesian,zhao2016learning,johnson2020revealing,pearce2025modeling}.
Most of them represent each ranker by one underlying component while a smaller literature allows continuous latent representation across several preference groups. However, even in these formulations, the primary goal is inference
on latent groups or ranking parameters rather than generative modeling of
the ranking population. In particular, it remains unclear how to learn a flexible
population distribution of ranker preferences from the observed rankings and use this
distribution to generate new rankers with previously unseen preference profiles.

Low-rank ranking models provide another flexible representation of heterogeneity by expressing ranker-specific utility $q_j$ through shared low-dimensional latent factors \citep{park2015preference,negahban2018learning,wu2018sql}. These models are primarily developed for utility estimation, preference completion, or ranking prediction. Moreover, the latent factors are generally identifiable only up to transformations and therefore do not define a canonical latent space on which a population distribution can be meaningfully learned and interpreted. A related limitation appears in classical generative rank models such as \citet{biernacki2013generative}, which restrict the ranking distribution to a specific low-dimensional parametric family.

Recent advances in generative AI have produced a wide range of flexible models,
including autoregressive architectures \citep{germain2015made,huang2025towards}, variational autoencoders \citep{jazbec2021scalable}, flows \citep{rezende2015variational,guo2025action},
and diffusion-based methods \citep{ho2020denoising}. While these approaches provide powerful general-purpose
tools for distribution learning and data generation, their direct application to ranking
data faces several difficulties. Modeling rankings directly requires learning a
distribution over a high-dimensional combinatorial space whose complexity grows rapidly
with the number of items. Introducing a generic neural latent representation can reduce
this dimensionality, but the resulting latent coordinates typically lack both rigorous finite-sample theoretical guarantees and statistically
identifiable interpretation for such a generative modeling of ranking data. Consequently, they do not directly address the joint statistical problem of
estimating heterogeneous ranker preferences, learning their population distribution, and
generating new rankings from that learned population.

These observations identify the gap addressed in this paper. Population-level ranking
generation requires more than either fitting a statistical ranking model or applying a
generic generative model. It requires a representation that is
low-dimensional enough to be estimated from one ranking per ranker, structured enough
to admit a statistical interpretation in terms of ranker preferences, and sufficiently
flexible to support distribution learning across a heterogeneous population.

\subsection{Notations}
Let $\mbb{Z}_{+}$ denote the set of all positive integers. For $n\in\mbb{Z}_{+}$, $[n]$ stands for the set $\{1,\ldots,n\}$ and $\mbf{1}_n$ denotes the $n$-dimensional vector of ones. For $k\in\mbb{Z}_{+}$, let $e_k$ be the Euclidean basis vector whose $k$-th entry is one and the rest entries are zero. For a matrix $A$, let $\|A\|$ be its operator norm, $\|A\|_F$ be its Frobenius norm, and $\|A\|_{\max}:=\max_{ij}|A_{ij}|$. For a vector $a$, use $\|a\|$ for its Euclidean norm and $\|a\|_{\infty}$ for its $\ell_\infty$ norm. We write $\mscr N(\mu,\Sigma)$ for the normal distribution with mean $\mu$ and covariance matrix $\Sigma$. For $a\in\mbb{R}$, let $[a]_{+}:=\max\{a,0\}$. For a vector, \([\cdot]_{+}\) is applied componentwise. For a set $A$, let $|A|$ denote its cardinality. We write $A\lesssim B$ (equivalently $B\gtrsim A$) if there exists a constant $C>0$ such that $A\le CB$. We write $A\asymp B$ if $A\lesssim B$ and $B\lesssim A$.

\section{Proposed method}\label{sec:ourmethod}

\subsection{Latent Preference Simplex Embedding (LPSE)}\label{sec:lpse}

We first recall the Plackett-Luce (PL) ranking model. For ranker
$j\in[m]$, let
\(
q_j=(q_{j1},\ldots,q_{jn})^\top\in\mathbb R^n
\)
denote the utility vector. PL model uses the following sequential softmax law for the probability of observing a top-$L_j$ ranking
\(
R_j=(i_{j1},\ldots,i_{jL_j})
\):
\begin{align}\label{eq:plmodel}
\mbb{P}(R_j\mid q_j)
=
\prod_{\ell=1}^{L_j}
\frac{\exp(q_{j i_{j\ell}})}
{\sum_{r\in A_{j\ell}}\exp(q_{jr})},
\end{align}
where
\(
A_{j\ell}
=
[n]\setminus\{i_{j1},\ldots,i_{j,\ell-1}\}
\)
is the remaining set before the $\ell$th selection. 

A fully unrestricted PL model considers unrelated utility vectors $\{q_j\}_{j=1}^m$. Directly building a generative model on these high-dimensional utility vectors is statistically challenging. More importantly, our goal is to construct a generative framework that respects the underlying ranking mechanism while retaining a clear statistical interpretation. To this end, the key observation is that empirical ranking data are heterogeneous in a structured way: different users may rank in substantially different ways while these
rankings often exhibit common preference types over the same collection of items. For example,
in movie rankings, some users may consistently place action movies near the top as one preference type,
whereas others may favor comedies or dramas; many users may also combine such preference types in a single ranking. This observation motivates us to represent each ranker's utility vector $q_j$ as a combination of latent utility vectors associated with some preference types so that each ranker aggregates their preference types when forming
a final ranking.

Formally, let \(h_1,\ldots,h_K\in\mathbb R^n\) denote latent utility
vectors associated with $K$ preference types for some $K\in\mbb{Z}_{+}$. For each $k\in [K]$, \(h_k=(h_{k1},\ldots,h_{kn})^\top\) records latent utilities of $n$ items for $k$-th preference type. Ranker \(j\) aggregates these
latent utilities to form an individual utility vector \(q_j\in
\mathbb R^n\):
\begin{align}\label{eq:lowrankq}
    q_j = \sum_{k=1}^K\pi_{jk}h_k,
\end{align}
where $\pi_{jk}$ is the weight of ranker $j$ on their preference type $k$. Let
\(
\pi_j=(\pi_{j1},\ldots,\pi_{jK})^\top\in\Delta^{K-1}:=
\left\{
x\in[0,1]^K:\sum_{k=1}^K x_k=1
\right\},
\)
where $\Delta^{K-1}$ is a $(K-1)$ dimensional latent preference simplex. Each vertex
\(h_k\) represents a pure preference type such as action movies, comedies,
and dramas in the above movie ranking example. To see how this simplex representation recovers the preference-type interpretation, consider two items \(a,b\in [n]\).
For any two items \(a\) and \(b\), let
\(
\mbb{P}_j(a\succ b)\) and \(
\mbb{P}_j(b\succ a)
\)
denote the probabilities that ranker \(j\) ranks \(a\) above \(b\) and \(b\)
above \(a\), respectively. Under the PL model \eqref{eq:plmodel}, the pairwise preference probability is a logistic function of the
utility difference \(q_{ja}-q_{jb}\):
\[
\log\frac{\mbb P_j(a\succ b)}{\mbb P_j(b\succ a)}
=
q_{ja}-q_{jb}\overset{\eqref{eq:lowrankq}}{=}
\sum_{k=1}^K\pi_{jk}(h_{ka}-h_{kb}).
\]
For preference type \(k\),
the utility difference \(h_{ka}-h_{kb}\) measures how strongly that type favors
\(a\) over \(b\): it is positive when the type tends to rank \(a\) above
\(b\), negative when it tends to rank \(b\) above \(a\), and larger in
magnitude when the preference is stronger. The weight \(\pi_{jk}\) reflects how much this utility difference under preference type $k$ contributes to ranker $j$'s tendency to rank item $a$ above item $b$.

LPSE is reminiscent of
topic models and latent-factor models in that ranker utilities are
represented as a collection of latent utilities. Its geometry, however,
is different. In topic models, each topic is typically a nonnegative probability
vector with a common unit-sum normalization while our latent utilities are signed vectors, and only their utility differences are relevant.

\subsection{Likelihood-based estimation of the latent preference simplex}\label{sec:MLE-VR}
\label{subsec:mle}
Let $Q = [q_1,\ldots,q_m]$, $H = [h_1,\ldots,h_K]$ and $\Pi=[\pi_1,\ldots,\pi_m]$ with $Q = H\Pi$. We first address the identifiability issue of the model parameters before we estimate $(\Pi,H)$.

The PL probabilities depend only on utility differences: for
any \(a\in\mathbb R^K\), define
\(
H^{(a)}
=
H+\mathbf 1_n a^\T.
\)
For ranker $j$ and item $i$,
\(
q_{ji}^{(a)}
=
\pi_j^\top h_i^{(a)}
=
\pi_j^\top h_i+\pi_j^\top a
=
q_{ji}+\pi_j^\top a.
\)
All utilities of ranker \(j\) are shifted by the
same constant such that
\(
\mbb{P}(R_j\mid\pi_j,H^{(a)})
=
\mbb{P}(R_j\mid\pi_j,H).
\)
It implies that the likelihood is consequently invariant to arbitrary row-wise shifts of
\(H\). We then impose a condition on $H$: (C1) \(
H^\T\mathbf 1_n=\mbf{0}.
\)
For the preference simplex parameter $(\Pi,H)$, we consider the conditions: (C2) For each preference type $k\in [K]$, there exists a pure ranker $j_k$ (i.e., $\pi_{j_k}=e_k$); (C3) $\{h_k\}_{k=1}^K$ are affinely independent. Under Conditions (C1)-(C3), all parameters are identifiable (up to a permutation if not pre-labeled and we will omit this permutation for notational convenience; see Proposition 2 in the supplementary material).
\begin{remark}
    Condition (C3) means \(
\sum_{k=1}^K \alpha_k h_k = \mbf{0}\ \text{and}\ 
\sum_{k=1}^K \alpha_k = 0
\)
imply all $\alpha_k$'s are zero. It ensures every utility vector $q_j$ admits a unique convex representation in \eqref{eq:lowrankq} by $h_1,\ldots,h_K$. This condition is weaker than $\rank(H)=K$ in density mixture model \citep{fan2026optimal} and DCMM network model \citep{jin2024mixed}. 
\end{remark}

\begin{remark}\label{rmk:identifiability}
Condition (C2) ensures that pure rankers serve as labeled anchors to
identify the latent preference simplex. This condition
is analogous to the pure-node condition in mixed-membership network
models \citep{jin2024mixed} and the anchor-region assumption in density mixtures \citep{fan2026optimal}. It can be relaxed to $\epsilon$-pure ranker, where the distance of some $\pi_{j_k}$ to $e_k$ is bounded by $\epsilon=o(1)$. There are some other identifiability conditions such as sufficiently scattered condition \citep{huang2019detecting} and volume-minimization identifiability \citep{he2023identifiability}. We choose Condition (C2) as it is simple and
statistically interpretable.
\end{remark}

In LPSE, each vertex is a column of the latent utility matrix $H$ and each column of the utility matrix $Q$ is a convex combination of all the vertices. This geometry motivates the use of vertex recovery on $\{q_j\}_{j=1}^m$ to recover the vertices $\{h_k\}_{k=1}^K$. Since $Q$ is unknown, we propose a constrained likelihood method to estimate $Q$ and then apply a vertex recovery algorithm to estimate $H$ and $\Pi$. Given the observation ranking data $\mcal{R}=\{R_1,\ldots,R_m\}$, the log-likelihood of the PL model in \eqref{eq:plmodel} is
\begin{align*}
\ell(\Pi,H)
&:=
\log \mbb{P}(\mathcal R\mid\Pi,H)=
\sum_{j=1}^m
\sum_{\ell=1}^{L_j}
\left[
\pi_j^\top h_{i_{j\ell}}
-
\log
\left[
\sum_{r\in A_{j\ell}}
\exp(\pi_j^\top h_r)
\right]
\right].
\end{align*}
We define
\(
\mcal{S}_{K,m}
=
\left\{
\Pi=[\pi_1,\ldots,\pi_m]\in\mathbb R^{K\times m}:
\pi_j\in\Delta^{K-1}
\text{ for every }j\in[m]
\right\}
\)
and
\(
\mathcal H_{n,K}
=
\left\{
H=[h_1,\ldots,h_K]\in\mathbb R^{n\times K}:
H^\T \mathbf 1_n=\mbf{0},\;
\|H\|_{\max}\le B
\right\}
\)
for some constant $B>0$, and consider the constrained maximum likelihood estimator
\begin{align}
\max_{\Pi\in\mcal S_{K,m},H\in\mcal H_{n,K}}\quad&
\sum_{j=1}^m
\sum_{\ell=1}^{L_j}
\left[
\pi_j^\top h_{i_{j\ell}}
-
\log
\left\{
\sum_{r\in A_{j\ell}}
\exp(\pi_j^\top h_r)
\right\}
\right].
\label{eq:constrained-mle-expanded}
\end{align}
Let $(\widetilde \Pi,\widetilde H)$ be a solution to \eqref{eq:constrained-mle-expanded} and $\widehat Q=\widetilde H \widetilde \Pi$. We also define 
\[
\mcal Q_{n,m,K}
:=
\left\{
Q=H\Pi:
H\in\mcal H_{n,K},\
\Pi\in\mcal S_{K,m}
\right\}.
\]
Since the log-likelihood depends on \((H,\Pi)\) only through their
product \(Q\), it is seen that \(\widehat Q\) is also an MLE of the utility matrix $Q$ with likelihood $\ell(Q)$ over the set \(\mcal Q_{n,m,K}\). However, $(\widetilde H, \widetilde \Pi)$ is not necessarily a consistent estimator of $(H,\Pi)$ since the factorization of \(Q\) is not uniquely determined by
the likelihood. Indeed, let \(A\in\mathbb{R}^{K\times K}\) be
invertible with
\(
A\mathbf 1_K=\mathbf 1_K.
\)
For any feasible pair \((\Pi,H)\), define
\(
\Pi'=A\Pi \) and \(
H'=HA^{-1}.
\)
Whenever \(\Pi'\in\mcal{S}_{K,m}\) and
\(H'\in\mcal{H}_{n,K}\), it always holds that
\(
H'\Pi'
=
(HA^{-1})(A\Pi)
=
H\Pi
\)
and 
\(
\ell(\Pi',H')=\ell(\Pi,H).
\)
The pair \((\widetilde\Pi,\widetilde H)\) is regarded only as a
preliminary factorization to obtain the estimator \(\widehat Q\).

To obtain an estimator for $(H,\Pi)$, we then apply a vertex recovery algorithm to
\(\widehat Q\) to obtain an estimator
\(\widehat H\). An estimator \(\widehat\Pi\) is subsequently
obtained from the barycentric coordinates of the columns of \(\widehat Q\)
relative to the estimated vertices as the columns of $\widehat H$. Here we
treat vertex recovery as a generic plug-in step in our method. Possible candidates could be Successive Projection (SP) \citep{araujo2001successive}, Sketched Vertex Search (SVS) and its variant SVS$^*$ \citep{jin2024mixed}. We pick SVS due to its additional denoising step. See more details in Section D.1.2 in the supplementary material. Operationally, SVS first applies a denoising step: given a tuning integer \(M_S\ge K\), 
partition the columns
\(\{\hat q_j\}_{j=1}^m\) into \(M_S\) clusters by \(k\)-means and
denote the resulting local centers by
\(\hat m_1,\ldots,\hat m_{M_S}\). Then, it selects \(K\) distinct
centers whose convex hull best approximates all \(L\) local centers:
\[
(\hat j_1,\ldots,\hat j_K)
\in
\argmin_{1\le j_1<\cdots<j_K\le M_S}
\max_{1\le \ell\le L}
\operatorname{dist}\!\left(
\hat m_\ell,\,
\operatorname{conv}
\{\hat m_{j_1},\ldots,\hat m_{j_K}\}
\right),
\]
and sets
\(
\widehat H
=
[\hat m_{\hat j_1},\ldots,
 \hat m_{\hat j_K}].
\)
After using SVS to estimate $H$, we propose to recover $\Pi$ by barycentric coordinates. Let
\(
\widehat D
=
[
\hat h_1-\hat h_K,\ldots,
\hat h_{K-1}-\hat h_K
].
\)
For each \(j\in[m]\), we first compute the  barycentric
coordinates
\(
\widetilde a_j
=
(\widehat D^\top\widehat D)^{-1}
\widehat D^\top
(\widehat q_j-\widehat h_K),
\)
and set
\(
\tilde\pi_j
=(
\tilde a_j^\T,
1-\mathbf 1_{K-1}^\top\tilde a_j)^\T.
\)
The final weight estimator is obtained by truncation and
renormalization:
\[
\hat\pi_j
=
\frac{[\tilde\pi_j]_+}
{\mathbf 1_K^\T[\tilde\pi_j]_+},
\qquad
\widehat\Pi
=
[\hat\pi_1,\ldots,\hat\pi_m],
\]
where for $a\in\mbb{R}$, $[a]_{+}:=\max\{a,0\}$ and for a vector, \([\cdot]{+}\) is applied componentwise.

\begin{remark}
One could enforce the pure ranker condition, or another
identifiability condition as stated in Remark \ref{rmk:identifiability} directly in the likelihood optimization.
However, such constraints are combinatorial or globally geometric and would
substantially complicate both theoretical analysis and computation of MLE with additional tuning and bias. Our method separates estimation from identification.
\end{remark}
\begin{remark}
The latent dimension \(K\) determines both the number of preference types and the dimension of the generative model trained in the next subsection. It may be selected using held-out PL likelihood/cross-validation (CV) \citep{stone1974cross,geisser1975predictive}, or information criteria \citep{mollica2017bayesian}. In the numerical studies, we select \(K\) using training-based model selection with an empirical screening. Details are provided in Section D.3 in the supplementary material.
\end{remark}

\subsection{Flow matching on the latent preference simplex}\label{sec:fm}
LPSE serves as the key representation on which our generative framework is built. Rather than directly modeling the highly
combinatorial ranking distribution (or through the \(n\)-dimensional utility vectors $\{q_j\}_{j=1}^m$), we
reduce the problem to learning and generating the unknown population distribution of the \(K\)-dimensional barycentric coordinates $\{\pi_j\}_{j=1}^m$. Formally, we regard the observed preference weights $\{\pi_j\}_{j=1}^m$ as one realization from an underlying population law \(\PP_\pi\) supported on the latent
preference simplex:
\(
\{\pi_j\}_{j=1}^m
\stackrel{\mathrm{i.i.d.}}{\sim}
\PP_\pi\), where \(
\PP_\pi\in\mathcal P(\Delta^{K-1}).
\)
Under this interpretation, the ranking distribution admits the hierarchical representation
\begin{align}\label{eq:hierPLmodel}
\pi_j
\sim
\PP_\pi,
\qquad
q_j=H\pi_j,
\qquad
R_j\mid\pi_j,H
\sim
\operatorname{PL}_{L_j}(H\pi_j),
\end{align}
where the $\text{PL}_{L_j}(\cdot)$ denotes the law given by the PL model \eqref{eq:plmodel}. This hierarchical representation motivates the following \textbf{L}atent \textbf{P}reference \textbf{S}implex \textbf{E}mbedding with \textbf{F}low \textbf{M}atching method (LPSE-FM, also see Algorithm \ref{alg:lpse-fm}): 
\begin{enumerate}
    \item Estimate the latent preference geometry
\((\widehat{\Pi},\widehat H)\) by constrained maximum likelihood in \eqref{eq:constrained-mle-expanded} and vertex recovery. Then hold \(\widehat H\) fixed and regard
$\{\hat\pi_j\}_{j=1}^m$ as noisy observations from the unknown
population \(\PP_\pi\).
\item We build a generative model $\widehat \PP_{\pi}\approx \PP_\pi$ and obtain  generated barycentric coordinates $\{\tilde{\pi}_j\}_{j\ge 1}\overset{\text{i.i.d.}}{\sim} \widehat \PP_\pi$.
\item Compute generated utility vectors $\tilde{q}_j=\widehat H\tilde{\pi}_j$ and generate the synthetic ranking by
\(
\widetilde R_j
\mid
\tilde\pi_j,\widehat H
\sim
\operatorname{PL}_{L_{\text{gen}}}(\widehat H\tilde\pi_j).
\)
\end{enumerate}
For step 2, we use flow matching to build a generative model $\widehat \PP_\pi$. Thus, the likelihood-based estimation in Section \ref{sec:MLE-VR} is formulated conditionally on the configuration $\Pi$, whereas $\PP_\pi$ provides the population-level model used to learn and generate out-of-sample preference weights.

Flow matching \citep{lipman2022flow} has recently emerged as one of the leading paradigms for deep generative modeling and achieved state-of-the-art performance across a wide range of generative modeling tasks. See \cite{lipman2024flow} for more details. Since our latent distribution $\PP_\pi$ lies in the simplex $\Delta^{K-1}$, we apply an additive log-ratio transformation of the simplex and then adopt a standard flow matching framework. Formally, define the log-ratio transformation
\begin{align}\label{eq:logratioT}
T:\Delta^{K-1}\to\mathbb R^{K-1},
\qquad
T(\pi)
=
\left(
\log\frac{\pi_1}{\pi_K},
\ldots,
\log\frac{\pi_{K-1}}{\pi_K}
\right)^\top.
\end{align}
Let the pushforward distribution be
\(
\PP_Z:=T_\#\PP_\pi=\PP_\pi\circ T^{-1}.
\)
We first fix \(\PP_0= \mscr N(0,I_{K-1})\) as a simple reference distribution. Let 
\(
Z_0\sim \PP_0
\)
and \(\PP_Z\) be our target distribution. Flow
matching aims to learn a time-dependent vector field, called the velocity field,
\(
v^*:\mathbb R^{K-1}\times[0,1]\rightarrow\mathbb R^{K-1}
\)
through a neural network estimator and considers solving the ODE
\begin{align}\label{eq:fmode}
\frac{dZ_t}{dt}
=
v^*(Z_t,t),
\qquad
Z_0\sim \PP_0,
\end{align}
such that the solution at $t=1$ satisfies $Z_1\sim \PP_Z$. There can be many such velocity fields. Flow matching proceeds by choosing a convenient path
\(
    Z_t = (1-t)Z_0+tZ_1\sim \PP_t,\ t\in [0,1].
\)
Along this path, the velocity field in \eqref{eq:fmode} is a conditional mean
\(
    v^*(x,t) = \mbb{E}[Z_1-Z_0\mid Z_t = x].
\)
To estimate it, given target observations
\(\{\hat z_j\}_{j=1}^m
\)
with $\hat z_j=T(\hat \pi_j)$ defined in \eqref{eq:logratioT}, the empirical loss becomes an MSE of regression:
\begin{align}\label{eq:lossneuralvector}
\widehat{\mathcal L}_{\mathrm{FM}}(\theta)
=
\frac{1}{m}
\sum_{j=1}^m
\mathbb E_{t\sim\text{Unif}[0,1],Z_0}
\left[
\left\|
v_\theta\!\left(
(1-t)Z_0+t\widehat z_j,t
\right)
-
(\widehat z_j-Z_0)
\right\|^2
\right].
\end{align}
We regress the neural velocity field $v_\theta\!\left(
(1-t)Z_0+t\hat z_j,t
\right)$ 
towards \(
\hat z_j-Z_0
\)
to obtain $\hat \theta$ and the learned velocity field $v_{\hat \theta}$. Let $\widehat\Phi_t(z_0)$ denote the exact solution at time $t$ to the
fitted flow-matching ODE
\begin{equation}\label{eq:fitted-fm-ode}
\frac{\mathrm d Z_t}{\mathrm dt}
=
v_{\hat\theta}(Z_t,t),
\qquad
Z_0=z_0,
\end{equation}
and write
\(
\widehat\Phi_t:
\mathbb R^{K-1}\to\mathbb R^{K-1}
\)
for the corresponding exact flow map. To solve \eqref{eq:fitted-fm-ode} numerically, we use the
explicit Euler scheme with $N$ equally spaced time steps. Let
the step size \(
h=N^{-1}\) for some $N\in\mbb{Z}_{+}$ and \(
t_k:=kh\) for \(
k=0,\ldots,N.
\)
For any initialization $z_0\in\mathbb R^{K-1}$, define
\(
\widehat\Phi_{0}^{(N)}(z_0):=z_0
\)
and, recursively,
\begin{equation}\label{eq:fm-euler}
\widehat\Phi_{t_{k+1}}^{(N)}(z_0)
=
\widehat\Phi_{t_k}^{(N)}(z_0)
+
h\,v_{\hat\theta}
\left(
\widehat\Phi_{t_k}^{(N)}(z_0),t_k
\right),
\qquad
k=0,\ldots,N-1.
\end{equation}
Thus, $\widehat\Phi_1^{(N)}$ is the numerical approximation of the exact flow map $\widehat\Phi_1$ in practice.  Then the flow matching distribution estimator of $\PP_Z$ is
\(
\widehat\PP_{Z,m}^{(N)}
:=
\left(\widehat\Phi_1^{(N)}\right)_\#\PP_0.
\)
Equivalently, if $Z_0\sim\PP_0$, then
\(
\widehat\Phi_1^{(N)}(Z_0)
\sim
\widehat\PP_{Z,m}^{(N)}.
\)
Finally, we estimate the
latent preference distribution $\PP_\pi$ and generate preference weights by $\{\tilde \pi_j\}_{j\ge 1}$:
\begin{align}\label{eq:learnedPpi-FM}
\widehat\PP_{\pi,m}^{(N)}
:=
(T^{-1})_\#\widehat\PP_Z^{(N)},\qquad \tilde\pi_j
=
T^{-1}\left(
\widehat\Phi_1^{(N)}(Z_{0,j})
\right),
\quad
j\geq1,
\end{align}
where
\(
Z_{0,1},Z_{0,2},\ldots
\stackrel{\mathrm{i.i.d.}}{\sim}
\PP_0
\)
such that
\(
\tilde\pi_1,\tilde\pi_2,\ldots
\stackrel{\mathrm{i.i.d.}}{\sim}
\widehat\PP_{\pi,m}^{(N)}
\)
and the inverse log-ratio map
\[
T^{-1}(z)
=
\frac{
(e^{z_1},\ldots,e^{z_{K-1}},1)^\top
}{
1+\sum_{k=1}^{K-1}e^{z_k}
},
\qquad
z=(z_1,\ldots,z_{K-1})^\top
\in\mathbb R^{K-1}.
\]

\begin{remark}
    Note that the log-ratio map $T$ is applied to $\hat \pi_{j}(k)/\hat \pi_{j}(K)$. For stability, when the MLE gives simplex coordinates close to zero, we can follow a common practice as in deep neural network training \citep{koloskova2023revisiting} to apply a 
clipping map
\(
\mathcal C_\tau(\hat \pi_j)(k)
=\max\{\hat \pi_j(k),\tau\}/
\sum_{k=1}^K\max\{\hat \pi_j(k),\tau\}\) for \(k\in[K]
\)
with a small threshold \(\tau>0\).
\end{remark}

\begin{algorithm}[H]
\caption{Latent Preference Simplex Embedding with Flow Matching (LPSE-FM)}
\label{alg:lpse-fm}
\begin{algorithmic}[1]

\State \textbf{Input:}
Observed rankings
\(\mathcal R\),
the number of preference types \(K\), the length and the number of rankings to be generated $L_{\text{gen}}$ and $m_{\mathrm{gen}}$, respectively.

\State \textbf{Latent preference estimation:}
Compute the constrained maximum likelihood estimator
\(
(\widetilde H,\widetilde\Pi)
\)
and the utility estimator
\(
\widehat Q=\widetilde H\widetilde\Pi.
\)

\State \textbf{Preference simplex recovery:}
Apply vertex recovery to \(\widehat Q\) to obtain the estimated latent utility matrix
\(
\widehat H
=
[\hat h_1,\ldots,\hat h_K],
\)
and the preference weight matrix
\(
\widehat\Pi
=
[\hat\pi_1,\ldots,\hat\pi_m].
\)

\State \textbf{Flow matching:}
Train a log-ratio-based flow matching on \(\{\hat\pi_j\}_{j=1}^m\) and obtain an estimated latent preference distribution
\(\widehat \PP_{\pi,m}^{(N)}\).

\State \textbf{Ranking generation:}
Generate
\(
\{\tilde\pi_j\}_{j=1}^{m_{\text{gen}}}
\overset{\mathrm{i.i.d.}}{\sim}
\widehat \PP_{\pi,m}^{(N)},
\)
set
\(
\tilde q_s=\widehat H\tilde\pi_s,
\)
and independently sample
\(
\widetilde R_s\mid \tilde\pi_s,\widehat H
\sim
\operatorname{PL}_{L_{\text{gen}}}(\tilde q_s),
\)
for $s\in [m_{\text{gen}}]$.

\State \textbf{Output:}
\(\{\widetilde R_s\}_{s=1}^{m_{\mathrm{gen}}}\overset{i.i.d.}{\sim}\widehat \PP_R\) and $\{\tilde\pi_j\}_{j=1}^{m_{\text{gen}}}$ (for interpretation).

\end{algorithmic}
\end{algorithm}

\begin{remark}
    Our latent distribution $\PP_\pi$ lies in the simplex $\Delta^{K-1}$. One alternative approach is to use recently developed manifold flow-matching methods \citep{chen2024flow,cheng2024categorical}. When applied to the probability simplex, however, these approaches typically endow the simplex with a
non-Euclidean geometry, for example, via the square-root embedding into
the positive orthant of the unit sphere, and construct flows along the
corresponding geodesics, leading to additional computational complexity.
\end{remark}

\section{Theoretical properties}
For any ranking length $\ell\in[n]$, define the population joint distribution
$\PP_{(R,\pi)}^{(\ell)}$ through
\(
\pi\sim\PP_\pi\),
\(R\mid\pi\sim\operatorname{PL}_{\ell}(H\pi),
\)
and let $\PP_R^{(\ell)}$ denote its marginal ranking distribution. Generally, for any probability measure $\PP_{\pi'}$ on the simplex $\Delta^{K-1}$ and
latent utility matrix $H'\in\mathbb{R}^{n\times K}$, define
$\PP_{(R',\pi')}^{(\ell)}$ through
\(
\pi'\sim\PP_{\pi'},\
R'\mid\pi'\sim\operatorname{PL}_{\ell}(H'\pi'),
\)
and let $\PP_{R'}^{(\ell)}$ denote the corresponding marginal ranking
distribution. In particular, taking
\(
\PP_{\pi'}=\widehat{\PP}_{\pi,m}^{(N)},\
H'=\widehat H\) and \(
R'=\widetilde R,
\)
gives the LPSE-FM generated ranking distribution
$\PP_{\widetilde R}^{(\ell)}$.

We first establish a result showing in the oracle case, our LPSE-FM reduces the generation from high-dimensional ranking data to the preference weights in the low-dimensional preference simplex. 

\begin{proposition}[Oracle KL reduction]\label{thm:oracle-kl-reduction}
If $H=H'$, then
\[
\KL
\!\left(
\PP_{R}^{(\ell)}
\,\middle\|\,
\PP_{R'}^{(\ell)}
\right)\le \KL
\!\left(
\PP_{(R,\pi)}^{(\ell)}
\,\middle\|\,
\PP_{(R',\pi')}^{(\ell)}
\right)
=
\KL
\!\left(
\PP_\pi
\,\middle\|\,
\PP_{\pi'}
\right),
\]
where $\KL(\cdot|| \cdot)$ denotes the Kullback-Leibler (KL) distance.
\end{proposition}

We next present the error rates when \((\Pi,H)\) is unknown. The analysis consists of two parts: the rate for the constrained MLE and the rate for vertex recovery.

\begin{theorem}[Rate of the restricted PL MLE]
\label{thm:restricted-pl-mle-rate}
Suppose $L_j\asymp L$ for $j\in [m]$. The restricted maximum likelihood estimator
\(\widehat Q\) satisfies
\begin{align*}
\frac{1}{mn}
\|\widehat Q-Q\|_F^2
=
O_{\mathbb P}
\left[
\left(
\frac{K}{L}
+
\frac{Kn}{mL}
\right)
\log(mnL)
\right].
\end{align*}
\end{theorem}

To accommodate a broad class of plug-in vertex recovery procedures, we introduce a generic stability condition; any procedure satisfying this condition can be used for our theoretical analysis. In particular, stability properties of this form are established in \cite{jin2024mixed} for the Sketched Vertex Search (SVS) and its variant SVS$^*$.
\begin{condition}
\label{cond:frobenius-vh}
For any noisy observation
\(
Q^\star
=
[q_1^\star,\ldots,q_m^\star]
\in\mathbb R^{n\times m},
\)
let
\(
H^\star
=
\VR(Q^\star)
\in\mathbb R^{n\times K}
\)
be the output of the vertex recovery. We say the vertex recovery is stable in Frobenius norm if there exists a constant \(C_{\mathrm{VH}}>0\) such that 
\[
\frac1n
\|H^\star-H\|_F^2
\le
C_{\mathrm{VH}}
\frac1{mn}
\|Q^\star-Q\|_F^2.
\]
\end{condition}

\begin{theorem}[Recovery of the LPSE]
\label{thm:simplex-weight-law}
Under the setting of Theorem \ref{thm:restricted-pl-mle-rate}, assume there exist constants
\(0<c_H\le C_H<\infty\) and
\(0<c_\Pi\le C_\Pi<\infty\) such that
\begin{align}\label{ass:gramHPi}
c_HI_K
\preceq
\frac1nH^\top H
\preceq
C_HI_K,\qquad 
c_\Pi I_K
\preceq
\frac1m\Pi\Pi^\top
\preceq
C_\Pi I_K,
\end{align}
where $A\preceq B$ means $B-A$ is positive semidefinite. Suppose the vertex recovery satisfies Condition~\ref{cond:frobenius-vh} and $K(m+n)\log(mnL)=o(mL)$. Then
\(\widehat D^\top\widehat D\) in the vertex recovery step is invertible with probability tending to one, and
\[
\frac1n
\|\widehat H-H\|_F^2+\frac1m
\|\widehat\Pi-\Pi\|_F^2
=
O_{\mathbb P}
\left[
\left(
\frac{K}{L}
+
\frac{Kn}{mL}
\right)
\log(mnL)
\right].
\]
\end{theorem}

\begin{remark}
     In the balance condition \eqref{ass:gramHPi}, the Gram matrix \(n^{-1}H^\top H\) controls the separation and
relative scale of the \(K\) latent preference types, while
\(m^{-1}\Pi\Pi^\top\) controls how broadly the rankers' weight vectors
span the corresponding preference directions. Together, they exclude
degenerate situations in which some preference types are nearly
indistinguishable or receive vanishingly little representation in the
population.
\end{remark}

Finally, we establish the error rate of our population-level generative modeling. Recall that $\mbb{P}_\pi$ denotes the probability measure from which $\{\pi_j\}_{j=1}^m$ are sampled and that $\widehat \PP_{\pi,m}^{(N)}$ denotes the learned distribution by flow matching defined in \eqref{eq:learnedPpi-FM}. Let
$\widetilde{\PP}_{\pi,m}^{(N)}$ denote the oracle flow matching distribution of $\widehat \PP_{\pi,m}^{(N)}$,
obtained by $\{\pi_j\}_{j=1}^m$ instead of $\{\hat \pi_j\}_{j=1}^m$. We decompose the generative modeling error measured by $1$-Wasserstein metric:
\begin{align*}
W_1\!\left(
\widehat{\PP}_{\pi,m}^{(N)},\PP_\pi
\right)
\le\;&
\underbrace{
W_1\!\left(
\widehat{\PP}_{\pi,m}^{(N)},
\widetilde{\PP}_{\pi,m}^{(N)}
\right)
}_{\text{estimation error}}
+
\underbrace{
W_1\!\left(
\widetilde{\PP}_{\pi,m}^{(N)},
\PP_\pi
\right)
}_{\text{flow matching error}} .
\end{align*}

\begin{theorem}[Wasserstein estimation error]
\label{lem:estimation-W1-simplex}
Under the setting of Theorem \ref{thm:simplex-weight-law}, we have:
\[
W_1\left(
\widehat{\PP}_{\pi,m}^{(N)},
\widetilde{\PP}_{\pi,m}^{(N)}
\right)
=
O_{\PP}\left[
\sqrt{
\left(
\frac{K}{L}
+
\frac{Kn}{mL}
\right)
\log(mnL)}
\right].
\]
\end{theorem}

For the flow matching error, we make the following standard assumptions \citep{zhou2025error,fukumizu2025flow,silveri2024theoretical,tsimpos2025optimal}.
\begin{assumption}[Lipschitz neural velocity]\label{ass:Lipvelo}
For some constants $L_v,B_v>0$, the neural velocity field $v_\theta$ in \eqref{eq:lossneuralvector} satisfies
\[
\|v_\theta(x,t)-v_\theta(y,s)\|
\le L_v(\|x-y\|+|t-s|),
\quad
\sup_{t\in[0,1]}\|v_\theta(0,t)\|\le B_v .
\]
\end{assumption}

\begin{assumption}[Oracle velocity estimation error]
\label{cond:velocity-estimation}
Let
$v_{\hat\theta}^{\mathrm{orc}}$
denote the oracle learned velocity field of $v_{\hat \theta}$ in \eqref{eq:lossneuralvector}, obtained by $\{\pi_j\}_{j=1}^m$ instead of $\{\hat \pi_j\}_{j=1}^m$. Then
\[
\int_0^1
\mathbb E_{Z\sim \PP_t}
\left[
\left\|
v_{\hat\theta}^{\mathrm{orc}}(Z,t)
-
v^*(Z,t)
\right\|^2
\right]dt
=
O_{\PP}\!\left(\varepsilon_{\mathrm{FM}}^2\right).
\]
\end{assumption}

\begin{assumption}[Moment condition on the latent preference distribution]
\label{cond:pi-log-moment}
\[
M_{2,\pi}
:=
\mathbb E_{\pi\sim\PP_\pi}
\left[
\|\log \pi\|^2
\right]
=
\mathbb E_{\pi\sim\PP_\pi}
\left[
\sum_{k=1}^K(\log \pi_k)^2
\right]
<\infty,
\]
where the logarithm is applied componentwise.
\end{assumption}
\begin{remark}
    Assumption \ref{ass:Lipvelo} is mild by enforcing norm-constrained neural networks and ensures  stability of the learned flow. Assumption \ref{cond:velocity-estimation} assumes $L_2$ consistency of the oracle learned velocity field as an estimator of the true velocity field. We will discuss the error $\varepsilon_{\text{FM}}$ later. For Assumption \ref{cond:pi-log-moment}, flow matching requires a finite second moment of the target distribution $\mbb{E}_{Z\sim\PP_Z}\|Z\|^2$ with $Z = T(\pi)$ and $T$ as the log-ratio map. The $\log$-moment condition on $\PP_\pi$ provides a simple sufficient condition for this requirement and is satisfied by common mixture distributions such as Dirichlet distribution and logistic-normal distributions.
\end{remark}

\begin{theorem}[Flow matching error]
\label{thm:fm-numerical-error}
Suppose Assumptions \ref{ass:Lipvelo},
\ref{cond:velocity-estimation}, and
\ref{cond:pi-log-moment} hold. Then
\[
W_1\left(
\widetilde{\PP}_{\pi,m}^{(N)},
\PP_\pi
\right)
=
O_{\mathbb P}\left(
\varepsilon_{\mathrm{FM}}+N^{-1}
\right).
\]
\end{theorem}

To present the final error rate of the proposed LPSE-FM method, we introduce the $1$-Wasserstein distance induced by the normalized Kemeny-Snell distance commonly used in the ranking literature. For a ranking $r\in\mathcal R_{n}$ and distinct items $a,b\in[n]$, we regard every ranked item as preceding every unranked item, while all unranked items are tied. Let $s_r(a,b)=1$ if $a$ is ranked above $b$ in $r$,
$s_r(a,b)=-1$ if $b$ is ranked above $a$, and $s_r(a,b)=0$ if both are unranked.
The normalized Kemeny-Snell distance is
\[
d_{\tau}^{(1/2)}(r,r')
=
\binom{n}{2}^{-1}\frac12
\sum_{1\leq a<b\leq n}
\left|s_r(a,b)-s_{r'}(a,b)\right|.
\]
For probability measures $\mu,\nu\in\mathcal P(\mathcal R_{n})$,
define
\[
W_{1,\tau}(\mu,\nu)
=
\inf_{\gamma\in\Gamma(\mu,\nu)}
\int d_{\tau}(r,r')\,d\gamma(r,r').
\]

\begin{theorem}[Wasserstein LPSE-FM error]
\label{thm:main}
Suppose the assumptions of Theorems
\ref{thm:simplex-weight-law},
\ref{lem:estimation-W1-simplex}, and
\ref{thm:fm-numerical-error} hold. Then
\begin{align*}
&W_1\left(
\PP_\pi,
\widehat{\PP}_{\pi,m}^{(N)}
\right)
+
\sup_{\ell\in [n]}W_{1,\tau}\left(
\PP_R^{(\ell)},
\PP_{\widetilde R}^{(\ell)}
\right)\\
&=
O_{\PP}\left[
\sqrt{
\left(
\frac{K}{L}
+
\frac{Kn}{mL}
\right)
\log(mnL)}
+
\varepsilon_{\mathrm{FM}}
+
N^{-1}
\right].
\end{align*}
Furthermore, if the step size of the  numerical ODE solver \eqref{eq:fm-euler} satisfies
\(
    N^{-1} = O\left[\sqrt{\left(
\frac{K}{L}
+
\frac{Kn}{mL}
\right)
\log(mnL)}+\varepsilon_{\mathrm{FM}}\right],
\)
then
\begin{align}
W_1\left(
\PP_\pi,
\widehat{\PP}_{\pi,m}^{(N)}
\right)
+
\sup_{\ell\in [n]}W_{1,\tau}\left(
\PP_R^{(\ell)},
\PP_{\widetilde R}^{(\ell)}
\right)=
O_{\PP}\left[
\sqrt{
\left(
\frac{K}{L}
+
\frac{Kn}{mL}
\right)
\log(mnL)}
+
\varepsilon_{\mathrm{FM}}
\right].
\label{eq:finalbound}
\end{align}
\end{theorem}

\begin{remark}
The final error rate \eqref{eq:finalbound} consists of two types of errors. The estimation error
\(
\sqrt{\left(
\frac{K}{L}
+
\frac{Kn}{mL}
\right)\log(mnL)}
\)
arises from replacing
$\{\pi_j\}_{j=1}^m$ by their estimates
$\{\hat{\pi}_j\}_{j=1}^m$.
The flow matching error $\varepsilon_{\text{FM}}$ depends on the sample size $m$, the smoothness $s>0$ of the density of the target distribution $\PP_\pi$ and the dimension $K-1$. We defer its detailed analysis to Section B in the supplementary material, where we show the flow matching error $=O_\PP\left( m^{-\frac{s}{2s+K-1}}\right)$.
\end{remark}

\section{Numerical experiments}\label{sec:experiment}

This section evaluates the proposed population-level generative framework on both synthetic and real ranking datasets. We compare our LPSE-FM method with two baselines: (i) Gaussian process variational autoencoders (GPVAE) \citep{jazbec2021scalable}, which learns a Gaussian process latent model by
maximizing the ELBO (evidence lower bound) using a variational posterior over the latent variables; after training, new observations are generated by sampling latent variables
from the fitted Gaussian process model and passing them through a decoder; (ii) Gaussian mixture variational autoencoders (GMVAE) \citep{lee2021meta}, which learns a Gaussian mixture latent model by maximizing the ELBO
with variational posteriors over both the mixture label and the continuous
latent variable; after training, new observations are generated by sampling a
latent variable from the fitted Gaussian mixture, and passing it through a decoder. For all methods, we evaluate the fidelity of the
generated ranking data through two metrics. 
\begin{enumerate}
    \item \textbf{Top-five frequency error.}
    For a collection $\mathcal A$ of rankings, let $f_i(\mathcal A)$ denote
    the proportion of rankings in $\mathcal A$ in which item $i$ appears
    among the first five positions. We define
    \begin{equation}
    \label{eq:simulation-delta-freq-k2}
        \Delta_{\mathrm{Top5}}
        =
        \left[
            \frac{1}{n}
            \sum_{i=1}^n
            \left(
                f_i(\widetilde{\mathcal R})-f_i(\mathcal R)
            \right)^2
        \right]^{1/2},
    \end{equation}
    where $\widetilde{\mathcal R}$ denotes the generated ranking sample.

    \item \textbf{Pairwise preference error.}
    For distinct items $a,b$, let $W_{ab}(\mathcal A)$ denote the proportion
    of rankings in $\mathcal A$ in which $a$ is ranked ahead of $b$. We define
    \begin{equation}
    \label{eq:simulation-delta-pair-k2}
        \Delta_{\mathrm{Pair}}
        =
        \left[
            \frac{1}{n(n-1)}
            \sum_{a\neq b}
            \left(
                W_{ab}(\widetilde{\mathcal R})
                -
                W_{ab}(\mathcal R)
            \right)^2
        \right]^{1/2}.
    \end{equation}
\end{enumerate}

We use the same batch size and the same number of iterations across all methods. All methods generate the same number of synthetic rankings for evaluation. The above metrics are reported by averaging over five independent runs. The code is available at \url{https://github.com/10Mzys/LPSE-FM}. Additional background and implementation details are deferred to Section D in the supplementary material. 

\subsection{Synthetic data}\label{sec:synthetic-data}
We generate complete rankings of
\(
n
\)
items from
\(
m
\)
rankers. Each ranking contains all $L=60$ items. We set
\(
K=3
\)
so that each ranker $j$ is represented by a preference weight
\(
\pi_j=(\pi_{j1},\pi_{j2},\pi_{j3})^\top.
\)
Let
\(
C_j\sim\operatorname{Categorical}(0.20,0.20,0.20,0.25,0.15).
\)
Define
\(
g_4(t)=
\bigl((1-c_t)t,\,(1-c_t)(1-t)\), \(c_t\bigr)^\top\), \(
c_t=0.18\sin^2(\pi t)
\)
and
\(
g_5(t)=
\bigl(
0.20+0.55t,\,
0.20+0.25\sin^2(\pi t),\,
0.60-0.55t-0.25\sin^2(\pi t)
\bigr)^\top.
\)
For $c=1,2,3$,
\[
\pi_j \mid C_j=c
\sim
\operatorname{Dirichlet}
\left(0.35\mathbf{1}_3+17.65e_c\right),
\]
and
\[
\pi_j=
\begin{cases}
\dfrac{g_4(T_j)+0.015\xi_j}
{\mathbf{1}^{\top}\!\left(g_4(T_j)+0.015\xi_j\right)},
& C_j=4,\\[10pt]
0.92
\dfrac{[g_5(T_j)]_+}
{\mathbf{1}^{\top}[g_5(T_j)]_+}
+0.08\zeta_j,
& C_j=5.
\end{cases}
\]
where
\[
T_j \mid C_j=c \sim
\begin{cases}
\operatorname{Beta}(0.7,0.7), & c=4,\\
\operatorname{Unif}(0,1), & c=5,
\end{cases}
\quad
\xi_j\sim\operatorname{Dirichlet}(\mathbf{1}_3),
\quad
\zeta_j\sim\operatorname{Dirichlet}(2\mathbf{1}_3).
\]
To construct \(H\in\mathbb R^{K\times n}\), partition
\(\{1,\ldots,n\}\) into \(K\) disjoint blocks
\(B_1,\ldots,B_K\) of approximately equal size. Let
\(\varepsilon_{ki}\overset{\mathrm{i.i.d.}}{\sim}\mscr N(0,0.25^2)\) and define
\[
H^{(0)}_{ki}
=
\varepsilon_{ki}
+
2.5\,\mathbf 1(i\in B_k)
-
\frac{2.5}{K-1}\mathbf 1(i\notin B_k),
\qquad
H_{ki}
=
H^{(0)}_{ki}
-
\frac1n\sum_{r=1}^n H^{(0)}_{kr}.
\]
Finally, a ranking
\(
R_j=(i_{j1},\ldots,i_{jL})
\)
is generated from the PL model \eqref{eq:hierPLmodel}.

We consider a setting $m\in \{600,1200\}$, $n\in \{60,120\}$ and generate $m$ rankings. The performance metrics are reported in Table~\ref{tab:simulation-results}. Overall, LPSE-FM consistently achieves the smallest $\Delta_{\mathrm{Top5}}$ error across all simulation settings, although the improvement over
GPVAE and GMVAE is relatively modest. However, LPSE-FM exhibits a substantially larger advantage in $\Delta_{\mathrm{Pair}}$ error. This indicates that it more accurately
captures the dependence structure among items rather than merely their
marginal frequencies. Since pairwise preference probabilities summarize the
relative ordering between every pair of items, this metric provides a much
more stringent assessment of the underlying ranking distribution. Additional experiments under other settings of $(m,n,K)$ are provided in Section E.1 in the supplementary material.

\begin{figure}[!t]
    \centering
    \includegraphics[width=0.5\linewidth]{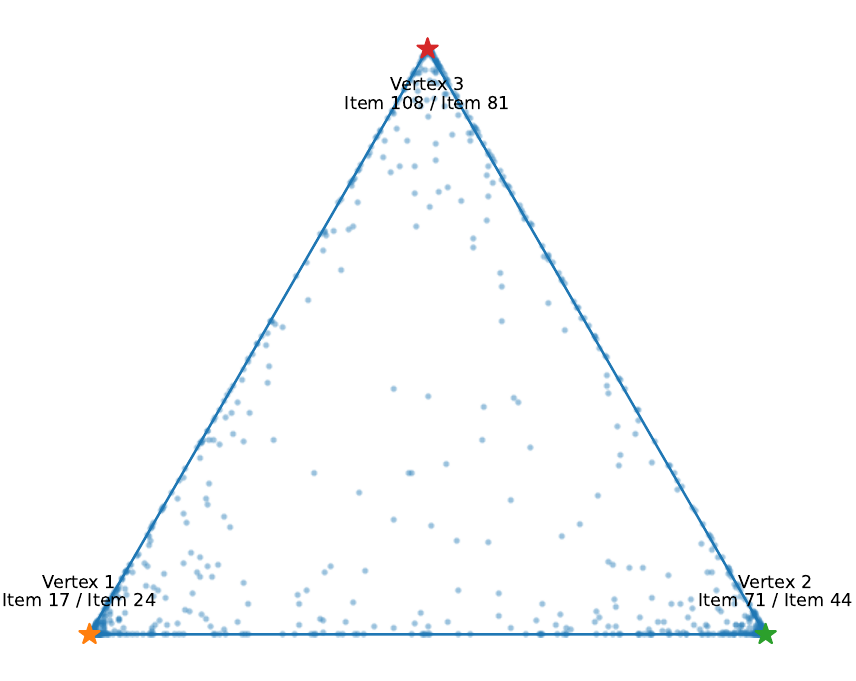}
    \caption{Generated preference weights on the estimated preference simplex by LPSE-FM under $(m,n)=(1200,120)$.}
    \label{fig:semantic-simplex}
\end{figure}

To show the interpretability of our LPSE-FM method, the generated preference weight
\(\tilde \pi_j\in\Delta^2\) can be visualized on a $2$D estimated preference simplex. Each point corresponds to one generated preference weight, while the three vertices represent the pure
preference types. To assign semantic meanings to the vertices, for vertex \(k\), we rank
the items according to \(\widehat H_{ki}\) and label the vertex using the two
items with the largest estimated utilities. In Figure \ref{fig:semantic-simplex}, the three
vertices are labeled by
\(\{\text{Item 17},\text{Item 24}\}\),
\(\{\text{Item 71},\text{Item 44}\}\), and
\(\{\text{Item 108},\text{Item 81}\}\) under $(m,n)=(1200,120)$. Points close to a vertex represent rankers dominated by one
preference type, and other points reflect contributions from mixed preference types. A complete list of
the leading items and their estimated utility scores are in 
Table \ref{tab:semantic_vertices}.

\begin{table}[!t]
\centering
\caption{Generative performance under different values of \(m\) and \(n\) measured by $\Delta_{\mathrm{Top5}}$ and $\Delta_{\mathrm{Pair}}$.}
\label{tab:simulation-results}
\scriptsize
\setlength{\tabcolsep}{7pt}
\renewcommand{\arraystretch}{1.80}
\begin{tabular}{l|cc|cc|cc|cc}
\hline
& \multicolumn{4}{c|}{\(m=600\)}
& \multicolumn{4}{c}{\(m=1200\)} \\
\hline
Method
& \multicolumn{2}{c|}{\(n=60\)}
& \multicolumn{2}{c|}{\(n=120\)}
& \multicolumn{2}{c|}{\(n=60\)}
& \multicolumn{2}{c}{\(n=120\)} \\
\hline
& \(\Delta_{\mathrm{Top5}}\)
& \(\Delta_{\mathrm{Pair}}\)
& \(\Delta_{\mathrm{Top5}}\)
& \(\Delta_{\mathrm{Pair}}\)
& \(\Delta_{\mathrm{Top5}}\)
& \(\Delta_{\mathrm{Pair}}\)
& \(\Delta_{\mathrm{Top5}}\)
& \(\Delta_{\mathrm{Pair}}\) \\
\hline
LPSE-FM
& \textbf{0.0155} & \textbf{0.0232}
& \textbf{0.0119} & \textbf{0.0458}
& \textbf{0.0111} & \textbf{0.0195}
& \textbf{0.0080} & \textbf{0.0388} \\

GPVAE
& 0.0186 & 0.0402
& 0.0162 & 0.0595
& 0.0113 & 0.0217
& 0.0143 & 0.0568 \\

GMVAE
& 0.0160 & 0.0314
& 0.0157 & 0.0614
& 0.0120 & 0.0227
& 0.0138 & 0.0561 \\
\hline
\end{tabular}
\end{table}

\begin{table}[!t]
\centering
\caption{Interpretation of the generated preference weights under $(m,n)=(1200,120)$. For each estimated vertex, the table reports its semantic label (defined by the two highest-utility items) and the six items having the largest estimated utilities and their utility scores.}
\label{tab:semantic_vertices}
\small
\setlength{\tabcolsep}{5pt}
\renewcommand{\arraystretch}{1.08}
\begin{tabular}{c c c c}
\hline
Vertex &
Semantic label &
Top item indices &
Top utility scores \\
\hline
1 &
(17, 24) &
17, 24, 9, 7, 6, 12 &
2.818, 2.807, 2.747, 2.736, 2.616, 2.582 \\

2 &
(71, 44) &
71, 44, 46, 64, 65, 74 &
2.875, 2.781, 2.763, 2.736, 2.640, 2.599 \\

3 &
(108, 81) &
108, 81, 86, 93, 119, 99 &
3.515, 3.260, 3.191, 3.073, 2.975, 2.898 \\
\hline
\end{tabular}
\end{table}

\subsection{Sushi preference dataset}\label{sec:sushi-experiment}
We evaluate LPSE-FM on the Sushi Preference Dataset
\citep{kamishima2003nantonac}, available at \url{https://www.kamishima.net/sushi/}. The dataset contains complete preference
rankings collected from 5000 individuals over a common set of ten sushi
types, using both five-point ratings and ranking responses. 

We randomly split the data into 4000
training rankings and 1000 testing rankings. For all methods, we generate $m_{\text{gen}}\in \{500,1000\}$ rankings. \(K=4\) is selected by minimizing the model-selection
criterion using the training data. Additional experiments under other settings of $m_{\text{gen}}$ are included in Section E.2 in the supplementary material.

\begin{figure}[!t]
    \centering
    \begin{minipage}[t]{0.48\linewidth}
        \centering
        \includegraphics[width=\linewidth]
        {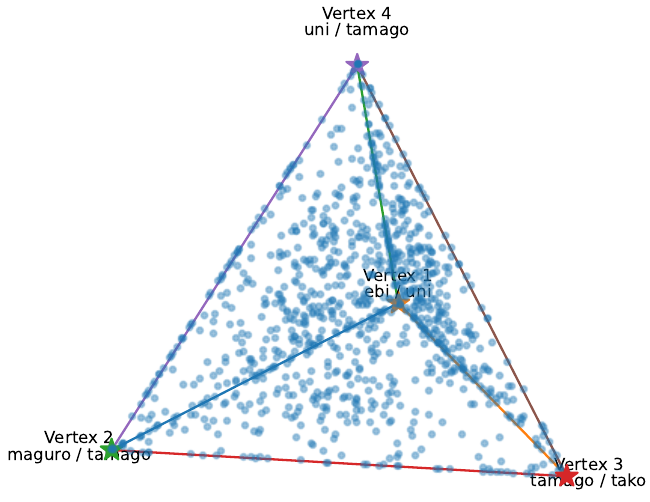}

        \small (a) Sushi
    \end{minipage}
    \hfill
    \begin{minipage}[t]{0.48\linewidth}
        \centering
        \includegraphics[width=\linewidth]
        {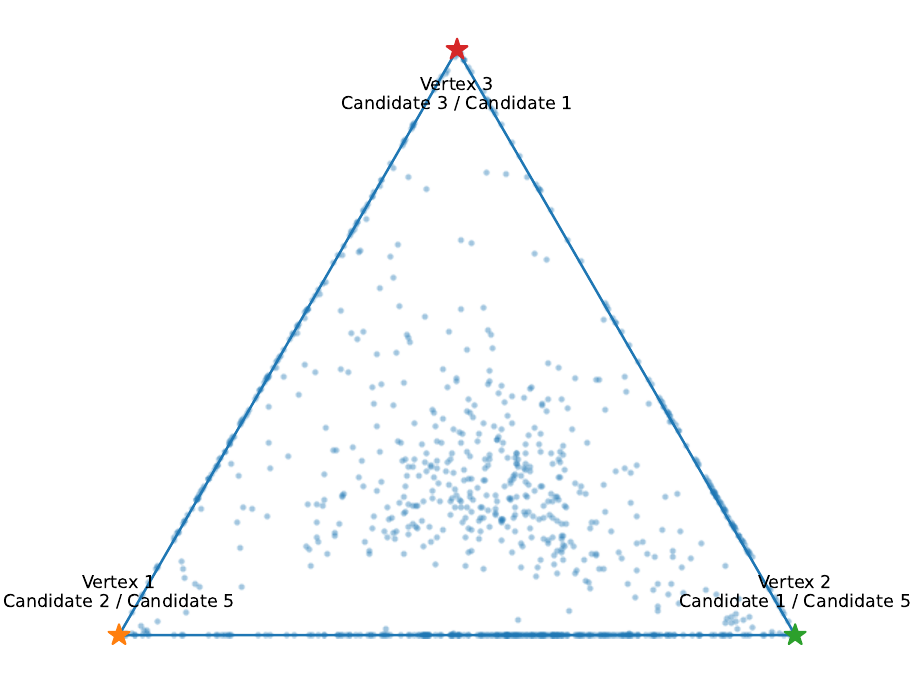}

        \small (b) APA election
    \end{minipage}
    \caption{Generated preference weights on the estimated preference
    simplices for (a) the Sushi preference dataset and (b) the APA
    presidential election dataset, with $m_{\mathrm{gen}}=1000$.}
    \label{fig:semantic-simplex-real}
\end{figure}

Table~\ref{tab:sushi-mgen-results} shows the performance metrics. Overall, LPSE-FM consistently achieves the smallest $\Delta_{\mathrm{Top5}}$ error and $\Delta_{\mathrm{Pair}}$ error under both settings. While GMVAE performs better than GPVAE and closer to LPSE-FM, our method achieves a substantially smaller $\Delta_{\mathrm{Pair}}$ error, indicating that
LPSE-FM more faithfully preserves the underlying dependence
structure among rankings rather than only marginal item frequencies. Also, the performance of all three methods changes only slightly as
\(m_{\mathrm{gen}}\) increases, suggesting that the
evaluation is stable with respect to the number of generated rankings. For interpretability, Figure~\ref{fig:semantic-simplex-real}(a) gives a substantive interpretation of the generated preference weights among the four
preference types of sushi under $m_{\text{gen}}=1000$. Vertex~1 is characterized
by \emph{ebi} (shrimp) and \emph{uni} (sea urchin), suggesting
a preference type towards distinctive seafood flavors. Vertex~2 is
characterized by \emph{maguro} (tuna) and \emph{tamago} (egg), representing a
more conventional type that combines a widely preferred fish item with a
mild, non-fish alternative. Vertex~3 emphasizes \emph{tamago} and \emph{tako}
(octopus), corresponding to a type that favors comparatively mild or
firm-textured items rather than rich fish varieties. Vertex~4 is characterized
by \emph{uni} and \emph{tamago}, reflecting a contrasting combination of the
rich, distinctive flavor of sea urchin and the mild sweetness of egg. Interestingly, \emph{tamago} appears among top two flavors in
three of the four types. This indicates that the preference
simplex separates consumers primarily through their secondary preferences
while identifying \emph{tamago} as a commonly favored item. Also, the broad interior dispersion in the simplex suggests
that sushi preferences are better represented by mixtures of
interpretable tastes than by four separated clusters. More details are provided in Table \ref{tab:sushi_semantic_vertices}. 

\begin{table}[!t]
\centering
\caption{Generative performance on the Sushi preference dataset for different numbers of generated rankings \(m_{\mathrm{gen}}\) measured by $\Delta_{\mathrm{Top5}}$ and $\Delta_{\mathrm{Pair}}$.}
\label{tab:sushi-mgen-results}
\small
\setlength{\tabcolsep}{7pt}
\renewcommand{\arraystretch}{1.00}
\begin{tabular}{l|cc|cc}
\hline
& \multicolumn{2}{c|}{\(m_{\mathrm{gen}}=500\)}
& \multicolumn{2}{c}{\(m_{\mathrm{gen}}=1000\)} \\
\hline
Method
& \(\Delta_{\mathrm{Top5}}\)
& \(\Delta_{\mathrm{Pair}}\)
& \(\Delta_{\mathrm{Top5}}\)
& \(\Delta_{\mathrm{Pair}}\) \\
\hline
LPSE-FM
& \textbf{0.03368}
& \textbf{0.04057}
& \textbf{0.03520}
& \textbf{0.04414} \\

GPVAE
& 0.12767
& 0.13683
& 0.12296
& 0.13277 \\

GMVAE
& 0.05955
& 0.06169
& 0.05641
& 0.05943 \\
\hline
\end{tabular}
\end{table}

\begin{table}[!t]
\centering
\caption{Interpretation of the generated preference weights for the Sushi preference dataset under $m_{\text{gen}}=1000$. For each vertex, the table reports its semantic label, the five sushi types with the largest estimated utilities, and the corresponding utility scores.}
\label{tab:sushi_semantic_vertices}

\small
\setlength{\tabcolsep}{4pt}
\renewcommand{\arraystretch}{1.00}

\resizebox{\linewidth}{!}{%
\begin{tabular}{c c l l}
\hline
Vertex &
Semantic label &
Top sushi &
Top utility scores \\
\hline

1 &
(ebi, uni) &
ebi, uni, anago, ika, ikura &
1.093, 0.913, 0.904, 0.564, 0.454 \\

2 &
(maguro, tamago) &
maguro, tamago, toro, ebi, ika &
2.817, 2.659, 1.199, 1.034, 0.784 \\

3 &
(tamago, tako) &
tamago, tako, maguro, anago, toro &
3.028, 2.602, 1.464, 0.672, 0.305 \\

4 &
(uni, tamago) &
uni, tamago, tako, maguro, ebi &
4.137, 3.666, 1.509, 0.686, -0.563 \\
\hline
\end{tabular}%
}
\end{table}

\subsection{PrefLib APA election data}\label{sec:apa-experiment}
We also consider the APA presidential election
dataset from the PrefLib preference repository
\citep{mattei2013preflib}, available at \url{https://preflib.github.io/PrefLib-Jekyll/dataset/00028}. It contains the results of the elections of the American Psychological Association (APA) between $1998$-$2009$. Voters submitted ordinal preferences over five candidates,
allowing the electorate to be represented through rank-order ballots. The APA
election data have consequently been used to study whether an organization
exhibits a coherent collective preference or instead displays the kinds of
aggregation difficulties highlighted in social-choice theory.

We use the \texttt{APA\_2000} profile distributed through PrefLib. The
original profile contains 20239 voters and five alternatives. Ballots are
stored in weighted form, with each distinct ranking accompanied by the number
of voters who submitted it; we expand these multiplicities to obtain
individual ranking observations. We retain the complete strict rankings so
that each observation is a permutation of the same five candidates. For this
common-alternative candidate structure, a generative model must
respect both the marginal tendency of candidates to appear near the top of
the ballot and the pairwise ordering frequencies across the electorate. We randomly partition the expanded rankings into 4000 training observations
and 1000 testing observations. LPSE-FM and both baseline methods are fitted
using only the training rankings and generate $m_{\text{gen}}\in \{500,1000\}$ rankings. \(K=3\) is selected by minimizing the model-selection
criterion using the training data. Since the dataset contains only five candidates, we replace $\Delta_{\mathrm{Top5}}$ error with $\Delta_{\mathrm{Top2}}$ error. Additional experiments under other settings of $m_{\text{gen}}$ are included in Section E.3 in the supplementary material.

Table~\ref{tab:apa-mgen-results} reports the generative performance. Interestingly, all methods achieve relatively similar $\Delta_{\mathrm{Top2}}$
errors, indicating that reproducing the marginal popularity of the most
preferred candidates is considerably less challenging. Nevertheless, LPSE-FM consistently achieves the smallest $\Delta_{\mathrm{Pair}}$
error, more faithfully capturing the common-alternative candidate structure. Although GPVAE presents a slightly smaller
$\Delta_{\mathrm{Top2}}$ error under \(m_{\mathrm{gen}}=1000\), its $\Delta_{\mathrm{Pair}}$ error
remains relatively larger than that of LPSE-FM, suggesting that matching
marginal frequencies alone does not guarantee an accurate representation of
the underlying preference distribution. Figure~\ref{fig:semantic-simplex-real}(b) shows the simplex interpretation with vertices labeled by the pairs Candidates \(\{2,5\}\),
Candidates \(\{1,5\}\), and
Candidates \(\{3,1\}\), respectively. It reveals a clear asymmetry that the generated preference weights concentrate toward the pair
Candidates \(\{1,5\}\), indicating that this
preference type is more prevalent than the other two. The relatively sparse
mass near Candidates \(\{2,5\}\) and
Candidates \(\{3,1\}\) vertices suggests a more specialized voting pattern. More details are provided in Table \ref{tab:apa_semantic_vertices}. 

\begin{table}[!t]
\centering
\caption{Generative performance on the APA presidential election
dataset for different numbers of generated rankings \(m_{\mathrm{gen}}\) measured by $\Delta_{\mathrm{Top2}}$ and $\Delta_{\mathrm{Pair}}$.}
\label{tab:apa-mgen-results}
\small
\setlength{\tabcolsep}{7pt}
\renewcommand{\arraystretch}{1.00}
\begin{tabular}{l|cc|cc}
\hline
& \multicolumn{2}{c|}{\(m_{\mathrm{gen}}=500\)}
& \multicolumn{2}{c}{\(m_{\mathrm{gen}}=1000\)} \\
\hline
Method
& \(\Delta_{\mathrm{Top2}}\)
& \(\Delta_{\mathrm{Pair}}\)
& \(\Delta_{\mathrm{Top2}}\)
& \(\Delta_{\mathrm{Pair}}\) \\
\hline
LPSE-FM
& \textbf{0.05602}
& \textbf{0.05200}
& 0.04288
& \textbf{0.03974} \\

GPVAE
& 0.05798
& 0.05641
& \textbf{0.04147}	
& 0.04886 \\

GMVAE
& 0.07245
& 0.06476
& 0.07169	
& 0.05959
 \\
\hline
\end{tabular}
\end{table}

\begin{table}[!t]
\centering
\caption{Interpretation of the generated preference weights for the APA presidential election dataset under $m_{\text{gen}}=1000$. For each vertex, the table reports its semantic label, the five candidates with the largest estimated utilities, and the corresponding utility scores.}
\label{tab:apa_semantic_vertices}
\small
\setlength{\tabcolsep}{4pt}
\renewcommand{\arraystretch}{0.95}
\begin{tabular}{c c l l}
\hline
Vertex &
Semantic label &
Top candidate indices &
Top utility scores \\
\hline
1 &
(2, 5) &
2, 5, 3, 4, 1 &
2.469, 2.034, 1.920, -1.515, -4.909 \\

2 &
(1, 5) &
1, 5, 2, 4, 3 &
2.952, 2.374, 1.164, -1.865, -4.626 \\

3 &
(3, 1) &
3, 1, 4, 5, 2 &
5.897, 3.552, 1.735, -5.450, -5.733 \\

\hline
\end{tabular}
\end{table}

Overall, the Sushi and APA experiments show that LPSE-FM can generate realistic
synthetic rankings while preserving interpretable population structure. In the Sushi application, the generated
rankings can facilitate consumer-preference and recommendation studies; in the
APA application, they can support the analysis and simulation of collective
choice and voting procedures.

\section{Discussion}
This paper develops a population-level generative framework for ranking data. The proposed LPSE-FM framework combines an interpretable low-dimensional latent preference simplex with flexible generative modeling, while retaining a probabilistic ranking mechanism for decoding latent preferences into rankings. The theoretical results show that ranking generation can be reduced to latent distribution learning and provide finite-sample guarantees that explicitly characterize the effects of the number of rankers, number of items, ranking length, and latent dimension. The numerical results further suggest that this structure is useful in practice: LPSE-FM consistently preserves population-level
fidelity, while the latent preference simplex provides a direct interpretation of population preference heterogeneity. 

\section*{Data availability}

All datasets used in this paper are publicly available from the sources
described in Section~\ref{sec:experiment}.

\bibliographystyle{abbrvnat}
\bibliography{reference}

\begin{thebibliography}{57}
\providecommand{\natexlab}[1]{#1}
\providecommand{\url}[1]{\texttt{#1}}
\expandafter\ifx\csname urlstyle\endcsname\relax
  \providecommand{\doi}[1]{doi: #1}\else
  \providecommand{\doi}{doi: \begingroup \urlstyle{rm}\Url}\fi

\bibitem[Ara{\'u}jo et~al.(2001)Ara{\'u}jo, Saldanha, Galvao, Yoneyama, Chame, and Visani]{araujo2001successive}
M.~C.~U. Ara{\'u}jo, T.~C.~B. Saldanha, R.~K.~H. Galvao, T.~Yoneyama, H.~C. Chame, and V.~Visani.
\newblock The successive projections algorithm for variable selection in spectroscopic multicomponent analysis.
\newblock \emph{Chemometrics and intelligent laboratory systems}, 57\penalty0 (2):\penalty0 65--73, 2001.

\bibitem[Biernacki and Jacques(2013)]{biernacki2013generative}
C.~Biernacki and J.~Jacques.
\newblock A generative model for rank data based on insertion sort algorithm.
\newblock \emph{Computational Statistics \& Data Analysis}, 58:\penalty0 162--176, 2013.

\bibitem[Caron et~al.(2014)Caron, Teh, and Murphy]{caron2014bayesian}
F.~Caron, Y.~W. Teh, and T.~B. Murphy.
\newblock Bayesian nonparametric plackett--luce models for the analysis of preferences for college degree programmes.
\newblock 2014.

\bibitem[Chen et~al.(2022)Chen, Gao, and Zhang]{chen2022optimal}
P.~Chen, C.~Gao, and A.~Y. Zhang.
\newblock Optimal full ranking from pairwise comparisons.
\newblock \emph{The Annals of Statistics}, 50\penalty0 (3):\penalty0 1775--1805, 2022.

\bibitem[Chen and Lipman(2024)]{chen2024flow}
R.~T. Chen and Y.~Lipman.
\newblock Flow matching on general geometries.
\newblock In \emph{International Conference on Learning Representations}, volume 2024, pages 47922--47945, 2024.

\bibitem[Chen et~al.(2019)Chen, Fan, Ma, and Wang]{chen2019spectral}
Y.~Chen, J.~Fan, C.~Ma, and K.~Wang.
\newblock Spectral method and regularized mle are both optimal for top-k ranking.
\newblock \emph{Annals of statistics}, 47\penalty0 (4):\penalty0 2204, 2019.

\bibitem[Chen et~al.(2024)Chen, Tan, Zhang, Yang, Sheng, Zhang, Wang, and Chua]{chen2024softmax}
Y.~Chen, J.~Tan, A.~Zhang, Z.~Yang, L.~Sheng, E.~Zhang, X.~Wang, and T.-S. Chua.
\newblock On softmax direct preference optimization for recommendation.
\newblock \emph{Advances in Neural Information Processing Systems}, 37:\penalty0 27463--27489, 2024.

\bibitem[Cheng et~al.(2024)Cheng, Li, Peng, and Liu]{cheng2024categorical}
C.~Cheng, J.~Li, J.~Peng, and G.~Liu.
\newblock Categorical flow matching on statistical manifolds.
\newblock \emph{Advances in Neural Information Processing Systems}, 37:\penalty0 54787--54819, 2024.

\bibitem[Chowdhury et~al.(2025)Chowdhury, Zick, and Allan]{chowdhury2025rankshap}
T.~Chowdhury, Y.~Zick, and J.~Allan.
\newblock Rankshap: Shapley value based feature attributions for learning to rank.
\newblock In \emph{International Conference on Learning Representations}, volume 2025, pages 36765--36794, 2025.

\bibitem[Dai et~al.(2021)Dai, Shen, Wang, and Qu]{dai2021scalable}
B.~Dai, X.~Shen, J.~Wang, and A.~Qu.
\newblock Scalable collaborative ranking for personalized prediction.
\newblock \emph{Journal of the American Statistical Association}, 116\penalty0 (535):\penalty0 1215--1223, 2021.

\bibitem[Dong et~al.(2026)Dong, Han, Jiang, and Xu]{dong2026statistical}
P.~Dong, R.~Han, B.~Jiang, and Y.~Xu.
\newblock Statistical ranking with dynamic covariates.
\newblock \emph{Journal of the Royal Statistical Society Series B: Statistical Methodology}, 88\penalty0 (1):\penalty0 221--238, 2026.

\bibitem[Duchi et~al.(2013)Duchi, Mackey, and Jordan]{duchi2013asymptotics}
J.~C. Duchi, L.~Mackey, and M.~I. Jordan.
\newblock The asymptotics of ranking algorithms.
\newblock \emph{The Annals of Statistics}, pages 2292--2323, 2013.

\bibitem[Fan et~al.(2025)Fan, Lou, Wang, and Yu]{fan2025ranking}
J.~Fan, Z.~Lou, W.~Wang, and M.~Yu.
\newblock Ranking inferences based on the top choice of multiway comparisons.
\newblock \emph{Journal of the American Statistical Association}, 120\penalty0 (549):\penalty0 237--250, 2025.

\bibitem[Fan et~al.(2026)Fan, Ke, and Shi]{fan2026optimal}
J.~Fan, Z.~T. Ke, and Z.~Shi.
\newblock Optimal demixing of nonparametric densities.
\newblock \emph{arXiv preprint arXiv:2603.27457}, 2026.

\bibitem[Fang et~al.(2024)Fang, Zhan, Ai, Mao, Su, Chen, and Liu]{fang2024scaling}
Y.~Fang, J.~Zhan, Q.~Ai, J.~Mao, W.~Su, J.~Chen, and Y.~Liu.
\newblock Scaling laws for dense retrieval.
\newblock In \emph{Proceedings of the 47th International ACM SIGIR Conference on Research and Development in Information Retrieval}, pages 1339--1349, 2024.

\bibitem[Fukumizu et~al.(2025)Fukumizu, Suzuki, Isobe, Oko, and Koyama]{fukumizu2025flow}
K.~Fukumizu, T.~Suzuki, N.~Isobe, K.~Oko, and M.~Koyama.
\newblock Flow matching achieves almost minimax optimal convergence.
\newblock In \emph{International Conference on Learning Representations}, volume 2025, pages 27608--27640, 2025.

\bibitem[Geisser(1975)]{geisser1975predictive}
S.~Geisser.
\newblock The predictive sample reuse method with applications.
\newblock \emph{Journal of the American statistical Association}, 70\penalty0 (350):\penalty0 320--328, 1975.

\bibitem[Germain et~al.(2015)Germain, Gregor, Murray, and Larochelle]{germain2015made}
M.~Germain, K.~Gregor, I.~Murray, and H.~Larochelle.
\newblock Made: Masked autoencoder for distribution estimation.
\newblock In \emph{International conference on machine learning}, pages 881--889. Pmlr, 2015.

\bibitem[Gormley and Murphy(2008)]{gormley2008mixture}
I.~C. Gormley and T.~B. Murphy.
\newblock A mixture of experts model for rank data with applications in election studies.
\newblock 2008.

\bibitem[Guo et~al.(2025)Guo, Xue, Huang, Wang, Wang, Wang, Wang, and Chen]{guo2025action}
H.~Guo, E.~Xue, L.~Huang, S.~Wang, X.~Wang, L.~Wang, J.~Wang, and S.~Chen.
\newblock Action is all you need: Dual-flow generative ranking network for recommendation.
\newblock \emph{arXiv preprint arXiv:2505.16752}, 2025.

\bibitem[Han and Xu(2025)]{han2025unified}
R.~Han and Y.~Xu.
\newblock A unified analysis of likelihood-based estimators in the plackett--luce model.
\newblock \emph{The Annals of Statistics}, 53\penalty0 (5):\penalty0 2077--2102, 2025.

\bibitem[He(2023)]{he2023identifiability}
S.~He.
\newblock \emph{Identifiability and estimation of mixed membership stochastic blockmodels}.
\newblock PhD thesis, University of Illinois at Urbana-Champaign, 2023.

\bibitem[Ho et~al.(2020)Ho, Jain, and Abbeel]{ho2020denoising}
J.~Ho, A.~Jain, and P.~Abbeel.
\newblock Denoising diffusion probabilistic models.
\newblock \emph{Advances in neural information processing systems}, 33:\penalty0 6840--6851, 2020.

\bibitem[Hosseini et~al.(2024)Hosseini, Mandal, and Puhan]{hosseini2024surprising}
H.~Hosseini, D.~Mandal, and A.~Puhan.
\newblock The surprising effectiveness of sp voting with partial preferences.
\newblock \emph{Advances in Neural Information Processing Systems}, 37:\penalty0 2787--2829, 2024.

\bibitem[Huang and Fu(2019)]{huang2019detecting}
K.~Huang and X.~Fu.
\newblock Detecting overlapping and correlated communities without pure nodes: Identifiability and algorithm.
\newblock In \emph{International Conference on Machine Learning}, pages 2859--2868. PMLR, 2019.

\bibitem[Huang et~al.(2025)Huang, Chen, Cao, Yang, Qi, Zhu, Han, Liu, Liu, Yao, et~al.]{huang2025towards}
Y.~Huang, Y.~Chen, X.~Cao, R.~Yang, M.~Qi, Y.~Zhu, Q.~Han, Y.~Liu, Z.~Liu, X.~Yao, et~al.
\newblock Towards large-scale generative ranking.
\newblock \emph{arXiv preprint arXiv:2505.04180}, 2025.

\bibitem[Jazbec et~al.(2021)Jazbec, Ashman, Fortuin, Pearce, Mandt, and R{\"a}tsch]{jazbec2021scalable}
M.~Jazbec, M.~Ashman, V.~Fortuin, M.~Pearce, S.~Mandt, and G.~R{\"a}tsch.
\newblock Scalable gaussian process variational autoencoders.
\newblock In \emph{International conference on artificial intelligence and statistics}, pages 3511--3519. PMLR, 2021.

\bibitem[Jin et~al.(2024)Jin, Ke, and Luo]{jin2024mixed}
J.~Jin, Z.~T. Ke, and S.~Luo.
\newblock Mixed membership estimation for social networks.
\newblock \emph{Journal of Econometrics}, 239\penalty0 (2):\penalty0 105369, 2024.

\bibitem[Johnson et~al.(2020)Johnson, Henderson, and Boys]{johnson2020revealing}
S.~R. Johnson, D.~A. Henderson, and R.~J. Boys.
\newblock Revealing subgroup structure in ranked data using a bayesian wand.
\newblock \emph{Journal of the American Statistical Association}, 115\penalty0 (532):\penalty0 1888--1901, 2020.

\bibitem[Kamishima(2003)]{kamishima2003nantonac}
T.~Kamishima.
\newblock Nantonac collaborative filtering: recommendation based on order responses.
\newblock In \emph{Proceedings of the ninth ACM SIGKDD international conference on Knowledge discovery and data mining}, pages 583--588, 2003.

\bibitem[Koloskova et~al.(2023)Koloskova, Hendrikx, and Stich]{koloskova2023revisiting}
A.~Koloskova, H.~Hendrikx, and S.~U. Stich.
\newblock Revisiting gradient clipping: Stochastic bias and tight convergence guarantees.
\newblock In \emph{International Conference on Machine Learning}, pages 17343--17363. PMLR, 2023.

\bibitem[Lee et~al.(2021)Lee, Min, Lee, and Hwang]{lee2021meta}
D.~B. Lee, D.~Min, S.~Lee, and S.~J. Hwang.
\newblock Meta-gmvae: Mixture of gaussian vae for unsupervised meta-learning.
\newblock \emph{ICLR}, 2:\penalty0 6, 2021.

\bibitem[Li et~al.(2025)Li, Tamkin, Goodman, and Andreas]{li2025eliciting}
B.~Li, A.~Tamkin, N.~Goodman, and J.~Andreas.
\newblock Eliciting human preferences with language models.
\newblock In \emph{International Conference on Learning Representations}, volume 2025, pages 80984--81013, 2025.

\bibitem[Lipman et~al.(2022)Lipman, Chen, Ben-Hamu, Nickel, and Le]{lipman2022flow}
Y.~Lipman, R.~T. Chen, H.~Ben-Hamu, M.~Nickel, and M.~Le.
\newblock Flow matching for generative modeling.
\newblock \emph{arXiv preprint arXiv:2210.02747}, 2022.

\bibitem[Lipman et~al.(2024)Lipman, Havasi, Holderrieth, Shaul, Le, Karrer, Chen, Lopez-Paz, Ben-Hamu, and Gat]{lipman2024flow}
Y.~Lipman, M.~Havasi, P.~Holderrieth, N.~Shaul, M.~Le, B.~Karrer, R.~T. Chen, D.~Lopez-Paz, H.~Ben-Hamu, and I.~Gat.
\newblock Flow matching guide and code.
\newblock \emph{arXiv preprint arXiv:2412.06264}, 2024.

\bibitem[Liu et~al.(2025)Liu, Zhang, Hu, Qian, and Chua]{liu2025preference}
S.~Liu, A.~Zhang, G.~Hu, H.~Qian, and T.-s. Chua.
\newblock Preference diffusion for recommendation.
\newblock In \emph{International Conference on Learning Representations}, volume 2025, pages 79844--79881, 2025.

\bibitem[Mao and Wu(2022)]{mao2022learning}
C.~Mao and Y.~Wu.
\newblock Learning mixtures of permutations: Groups of pairwise comparisons and combinatorial method of moments.
\newblock \emph{The Annals of Statistics}, 50\penalty0 (4):\penalty0 2231--2255, 2022.

\bibitem[Mattei and Walsh(2013)]{mattei2013preflib}
N.~Mattei and T.~Walsh.
\newblock Preflib: A library for preferences http://www. preflib. org.
\newblock In \emph{International conference on algorithmic decision theory}, pages 259--270. Springer, 2013.

\bibitem[Mollica and Tardella(2017)]{mollica2017bayesian}
C.~Mollica and L.~Tardella.
\newblock Bayesian plackett--luce mixture models for partially ranked data.
\newblock \emph{Psychometrika}, 82\penalty0 (2):\penalty0 442--458, 2017.

\bibitem[Mukherjee et~al.(2024)Mukherjee, Lalitha, Kalantari, Deshmukh, Liu, Ma, and Kveton]{mukherjee2024optimal}
S.~Mukherjee, A.~Lalitha, K.~Kalantari, A.~Deshmukh, G.~Liu, Y.~Ma, and B.~Kveton.
\newblock Optimal design for human preference elicitation.
\newblock \emph{Advances in Neural Information Processing Systems}, 37:\penalty0 90132--90159, 2024.

\bibitem[Murphy and Martin(2003)]{murphy2003mixtures}
T.~B. Murphy and D.~Martin.
\newblock Mixtures of distance-based models for ranking data.
\newblock \emph{Computational statistics \& data analysis}, 41\penalty0 (3-4):\penalty0 645--655, 2003.

\bibitem[Negahban et~al.(2018)Negahban, Oh, Thekumparampil, and Xu]{negahban2018learning}
S.~Negahban, S.~Oh, K.~K. Thekumparampil, and J.~Xu.
\newblock Learning from comparisons and choices.
\newblock \emph{Journal of Machine Learning Research}, 19\penalty0 (40):\penalty0 1--95, 2018.

\bibitem[Park et~al.(2015)Park, Neeman, Zhang, Sanghavi, and Dhillon]{park2015preference}
D.~Park, J.~Neeman, J.~Zhang, S.~Sanghavi, and I.~Dhillon.
\newblock Preference completion: Large-scale collaborative ranking from pairwise comparisons.
\newblock In \emph{International Conference on Machine Learning}, pages 1907--1916. PMLR, 2015.

\bibitem[Pearce and Erosheva(2025)]{pearce2025modeling}
M.~Pearce and E.~A. Erosheva.
\newblock Modeling preferences: A bayesian mixture of finite mixtures for rankings and ratings.
\newblock \emph{Journal of the American Statistical Association}, 120\penalty0 (551):\penalty0 1621--1632, 2025.

\bibitem[Qiu et~al.(2025)Qiu, Wang, Zhang, Zheng, Zhu, Fan, Zhang, Wang, and Wang]{qiu2025unirom}
J.~Qiu, Z.~Wang, F.~Zhang, Z.~Zheng, J.~Zhu, J.~Fan, T.~Zhang, H.~Wang, and X.~Wang.
\newblock Unirom: Unifying online advertising ranking as one model.
\newblock In \emph{Proceedings of the 34th ACM International Conference on Information and Knowledge Management}, pages 2440--2449, 2025.

\bibitem[Rezende and Mohamed(2015)]{rezende2015variational}
D.~Rezende and S.~Mohamed.
\newblock Variational inference with normalizing flows.
\newblock In \emph{International conference on machine learning}, pages 1530--1538. PMLR, 2015.

\bibitem[Shi et~al.(2025)Shi, Liu, Long, Su, and Xiao]{shi2025fundamental}
Z.~Shi, K.~Liu, Q.~Long, W.~J. Su, and J.~Xiao.
\newblock Fundamental limits of game-theoretic llm alignment: Smith consistency and preference matching.
\newblock \emph{arXiv preprint arXiv:2505.20627}, 2025.

\bibitem[Silveri et~al.(2024)Silveri, Conforti, and Durmus]{silveri2024theoretical}
M.~G. Silveri, G.~Conforti, and A.~Durmus.
\newblock Theoretical guarantees in kl for diffusion flow matching.
\newblock \emph{Advances in Neural Information Processing Systems}, 37:\penalty0 138432--138473, 2024.

\bibitem[Stone(1974)]{stone1974cross}
M.~Stone.
\newblock Cross-validatory choice and assessment of statistical predictions.
\newblock \emph{Journal of the royal statistical society: Series B (Methodological)}, 36\penalty0 (2):\penalty0 111--133, 1974.

\bibitem[Tsimpos et~al.(2025)Tsimpos, Ren, Zech, and Marzouk]{tsimpos2025optimal}
P.~Tsimpos, Z.~Ren, J.~Zech, and Y.~Marzouk.
\newblock Optimal scheduling of dynamic transport.
\newblock \emph{arXiv preprint arXiv:2504.14425}, 2025.

\bibitem[Wu et~al.(2018)Wu, Hsieh, and Sharpnack]{wu2018sql}
L.~Wu, C.-J. Hsieh, and J.~Sharpnack.
\newblock Sql-rank: A listwise approach to collaborative ranking.
\newblock In \emph{International Conference on Machine Learning}, pages 5315--5324. PMLR, 2018.

\bibitem[Xiao et~al.(2025)Xiao, Li, Xie, Getzen, Fang, Long, and Su]{xiao2025algorithmic}
J.~Xiao, Z.~Li, X.~Xie, E.~Getzen, C.~Fang, Q.~Long, and W.~J. Su.
\newblock On the algorithmic bias of aligning large language models with rlhf: Preference collapse and matching regularization.
\newblock \emph{Journal of the American Statistical Association}, 120\penalty0 (552):\penalty0 2154--2164, 2025.

\bibitem[Yang et~al.(2024)Yang, Sang, Wang, Chen, Wang, He, Peng, Lin, Gan, and Shao]{yang2024parallel}
Z.~Yang, L.~Sang, H.~Wang, W.~Chen, L.~Wang, J.~He, C.~Peng, Z.~Lin, C.~Gan, and J.~Shao.
\newblock Parallel ranking of ads and creatives in real-time advertising systems.
\newblock In \emph{Proceedings of the AAAI Conference on Artificial Intelligence}, volume~38, pages 9278--9286, 2024.

\bibitem[Yin et~al.()Yin, Boughanmi, and Mukherjee]{yinexpress}
M.~Yin, K.~Boughanmi, and A.~Mukherjee.
\newblock Express: Modeling dynamic consumer preferences from few-shot data: A meta-learning approach.
\newblock \emph{Journal of Marketing Research}, page 00222437261458290.

\bibitem[Zhao et~al.(2016)Zhao, Piech, and Xia]{zhao2016learning}
Z.~Zhao, P.~Piech, and L.~Xia.
\newblock Learning mixtures of plackett-luce models.
\newblock In \emph{International Conference on Machine Learning}, pages 2906--2914. PMLR, 2016.

\bibitem[Zhou and Liu(2025)]{zhou2025error}
Z.~Zhou and W.~Liu.
\newblock An error analysis of flow matching for deep generative modeling.
\newblock In \emph{Forty-second International Conference on Machine Learning}, 2025.

\bibitem[Zhu et~al.(2023)Zhu, Jiang, Liu, and Deng]{zhu2023partition}
W.~Zhu, Y.~Jiang, J.~S. Liu, and K.~Deng.
\newblock Partition--mallows model and its inference for rank aggregation.
\newblock \emph{Journal of the American Statistical Association}, 118\penalty0 (541):\penalty0 343--359, 2023.

\end{thebibliography}

\end{document}